\documentclass[journal,comsoc,10pt,twocolumn]{IEEEtran}
\usepackage[T1]{fontenc}
\usepackage{amsmath}
\usepackage[cmintegrals]{newtxmath}

\usepackage{color}
\usepackage{colortbl}

\usepackage{cite}
\usepackage{enumitem}

\usepackage{url}
\usepackage{booktabs}
\usepackage{enumitem}
\usepackage{tabularx}
\newcolumntype{L}[1]{>{\raggedright\arraybackslash}p{#1}}
\newcolumntype{C}[1]{>{\centering\arraybackslash}p{#1}}
\newcolumntype{R}[1]{>{\raggedleft\arraybackslash}p{#1}}

\usepackage[binary-units = true, range-phrase=--, per-mode=symbol]{siunitx}
\DeclareSIUnit{\decibelm}{dBm}
\DeclareSIUnit{\Joule}{Joule}

\ifCLASSINFOpdf
   \usepackage[pdftex]{graphicx}
\else
\fi

\begin{document}

\title{Time Synchronization in 5G Wireless Edge: Requirements and Solutions for Critical-MTC}

\author{Aamir~Mahmood,
        Muhammad~Ikram~Ashraf,
				Mikael Gidlund,
        Johan~Torsner,
				and~Joachim Sachs% <-this % stops a space
\thanks{A. Mahmood and M. Gidlund are with Mid Sweden University, Sweden e-mail: aamir.mahmood@miun.se}% <-this % stops a space
\thanks{M. I. Ashraf, J. Torsner and J. Sachs are with Ericsson Research.}% <-this % stops a space
%\thanks{Manuscript received April 19, 2005; revised August 26, 2015.}
\vspace{-20pt}}

\maketitle

\begin{abstract}

Wireless edge is about distributing intelligence to the wireless devices 
wherein the distribution of accurate time reference is essential for 
time-critical machine-type communication (cMTC). In 5G-based cMTC, enabling time 
synchronization in the wireless edge means moving beyond the current 
synchronization needs and solutions in 5G radio access. In this article, we analyze the 
device-level synchronization needs of potential cMTC applications:
industrial automation, power distribution, vehicular communication, and live 
audio/video production. We present an over-the-air (OTA) synchronization 
scheme comprised of 5G air interface parameters, and discuss their associated 
timing errors. We evaluate the estimation error in device-to-base station 
propagation delay from timing advance (TA) under random errors and show how 
to reduce the estimation error. In the end, we identify the random errors 
specific to dense multipath fading environments and discuss countermeasures. 
\end{abstract}

% Note that keywords are not normally used for peerreview papers.

\IEEEpeerreviewmaketitle

%%%%%%%%%%%%%%%%%%%%%%%%%%%%%%%%%%%%%%%%%%%%%%%%%%%%%%%%%%%%%%%%%%%%%
%%%%%%%%%%%%%%%%%%%%%%%%%%%%%%%%%%%%%%%%%%%%%%%%%%%%%%%%%%%%%%%%%%%%%
%%%%%%%%%%%%%%%%%%%%%%%%%%%%%%%%%%%%%%%%%%%%%%%%%%%%%%%%%%%%%%%%%%%%%
%%%%%%%%%%%%%%%%%%%%%%%%%%%%%%%%%%%%%%%%%%%%%%%%%%%%%%%%%%%%%%%%%%%%%
%%%%%%%%%%%%%%%%%%%%%%%%%%%%%%%%%%%%%%%%%%%%%%%%%%%%%%%%%%%%%%%%%%%%%
%Time synchronization is critical for the operation of distributed systems in networked environments.%%%%%%%%%%%%%%%%%%%%%%%%%%%%%%%%%%%%%%%%%%%%%%%%%%%%%%%%%%%%%%%%%%%%
%%%%%%%%%%%%%%%%%%%%%%%%%%%%%%%%%%%%%%%%%%%%%%%%%%%%%%%%%%%%%%%%%%%%%
%%%%%%%%%%%%%%%%%%%%%%%%%%%%%%%%%%%%%%%%%%%%%%%%%%%%%%%%%%%%%%%%%%%%%
%%%%%%%%%%%%%%%%%%%%%%%%%%%%%%%%%%%%%%%%%%%%%%%%%%%%%%%%%%%%%%%%%%%%%
%%%%%%%%%%%%%%%%%%%%%%%%%%%%%%%%%%%%%%%%%%%%%%%%%%%%%%%%%%%%%%%%%%%%%
%%%%%%%%%%%%%%%%%%%%%%%%%%%%%%%%%%%%%%%%%%%%%%%%%%%%%%%%%%%%%%%%%%%%%
%%%%%%%%%%%%%%%%%%%%%%%%%%%%%%%%%%%%%%%%%%%%%%%%%%%%%%%%%%%%%%%%%%%%%
\section{Introduction} 
\IEEEPARstart{T}{he} very vision of 
Industry 4.0---making the industrial processes intelligent, efficient, and 
safer---is tied to real-time automation and control of dynamic industrial 
systems over wireless 
networks. However, what is achievable by existing wireless networking 
solutions in terms of communication reliability and latency is not sufficient 
for critical machine-type communication (cMTC). Instead, ultra-reliable and 
low-latency communication (URLLC) is required. Yet, the key sectors in 
cMTC---factory automation, power distribution, vehicular communication, live audio/video production, etc.~\cite{3gpp.22.804}---require precise time synchronization up to device level. 
If we take discrete manufacturing as an example, devices require synchronized coordination for timely/sequential execution of tasks such as 
assembly, picking, welding, and palletizing. In power distribution, monitoring 
and fault localization require perfectly synchronized measurement 
units. Hence in cMTC applications, where ultra-reliability is vital for 
the safety of processes, equipment and users, and low latency for real-time 
functionality of applications, time synchronization is intrinsic to real-time 
coordination and interaction among devices. 

Supporting determinism, in terms of reliability of $10^{-5}--10^{-6}$ 
and latency up to \SI{1}{\milli\second}, and synchronism with jitter below
\SI{1}{\micro\second} is the focus of 3GPP uses cases within 5G URLLC.
These requirements add challenges to the 
design of the radio access network (RAN) from the physical layer leading up to 
the radio resource control (RRC) layer. 3GPP Rel-15 and 16 specify new 
features for New Radio (NR) including
%In 3GPP release 
%15\&16 specifications for New Radio (NR), the newly proposed 
%features include 
faster scheduling, short and 
robust transmissions, repetitions, faster retransmissions, preemption, and 
packet duplication as well as multiconnectivity architectures~\cite{Sachs}. 
However, the importance of time synchronization, which is a crucial 
component of the new RAN technologies and services, moves beyond RAN to 
provide an accurate time instant on a common time base to the devices and is 
perceived as a key enabler for cMTC applications. 
%\textcolor{red}{Typically, 
%time synchronization can be achieved by using the global positioning system 
%(GPS), which can provide the desired accuracy for industrial networks. However, this 
%solution is fairly expensive, and impractical in indoor deployments.} 
The devices can achieve perfect time alignment to the coordinated universal time (UTC) by employing a 
global positioning system (GPS) receiver; however, such a solution is expensive and 
mainly impractical for indoor industrial deployments.

In industrial measurement and control systems, time is synchronized 
separately from the data flow~\cite{TSN_Finn}. Since many application require synchronization, it is an essential part of the 
Ethernet-based networks such as PROFINET and the rapidly evolving open standard: 
time-sensitive networking (TSN) by IEEE~802.1 task group~\cite{TSN2}. TSN defines a 
series of standards to support URLLC use cases over best-effort Ethernet 
networks. The synchronization procedures for industrial 
systems are rooted in precision time protocol (PTP), defined by the IEEE~1588 standard. All the Ethernet-based automation 
networks utilize IEEE~1588 variants, known as PTP profiles. In TSN, for instance, 
the transport of precise timing and synchronization is performed using 
the IEEE~802.1AS standard.

\begin{figure*}[!t] 	
	\centering 		
	\includegraphics[width=0.80\linewidth]{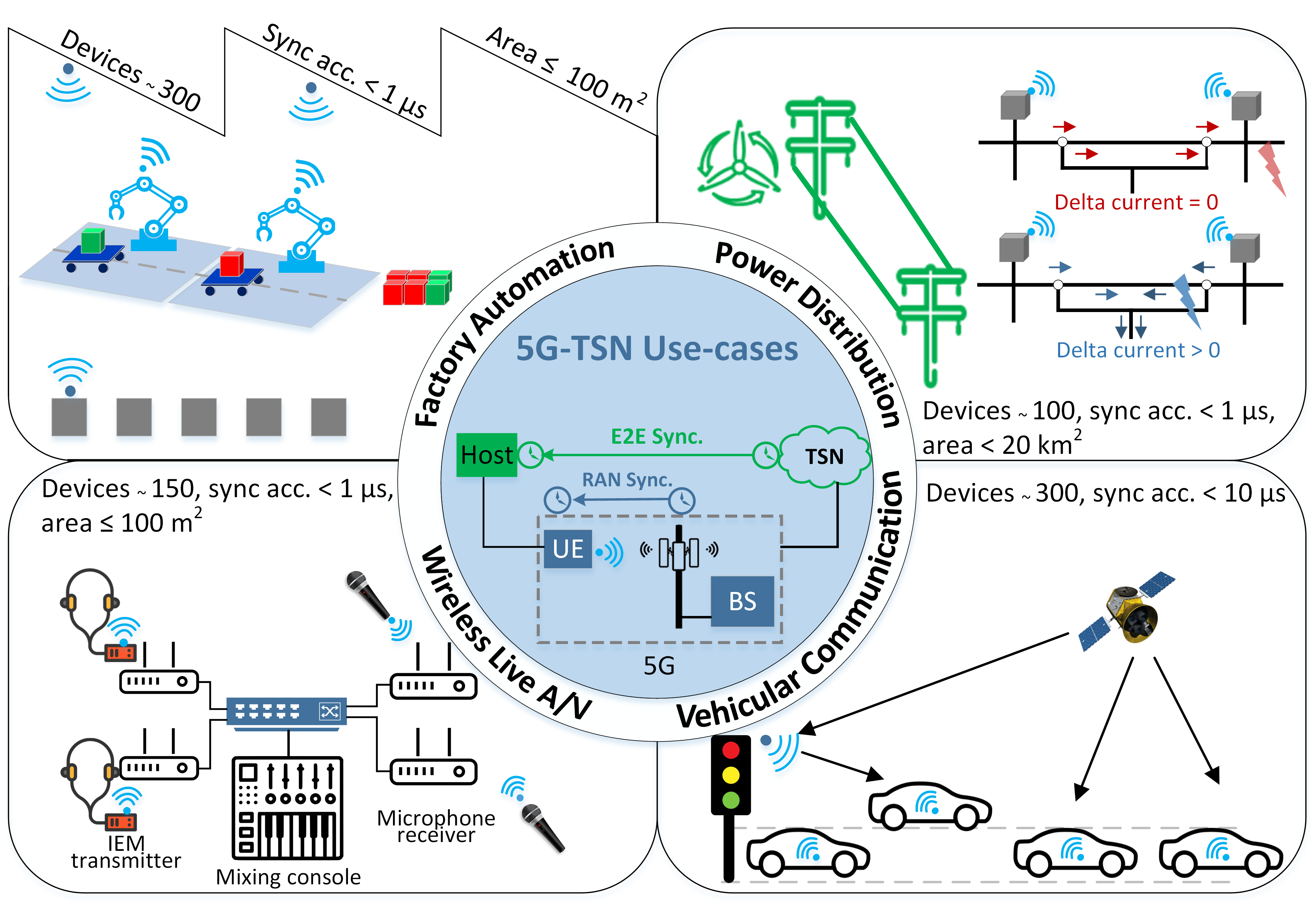}
 	\caption{Critical machine type communication (cMTC) use cases requiring 
device-level time synchronization and an enabling solution over 5G-TSN network.} 	
	\label{fig:usecases} 
	\vspace{-10pt}
\end{figure*}

Establishing over-the-air (OTA) accurate time reference at device level requires transfer/exchange of timestamps between controller (i.e., BS) and devices.
%which is beyond current frame alignment 
%procedures used for relative alignment of uplink/downlink transmissions.
Additionally, 5G is expected to coexist with existing industrial Ethernet and 
emerging TSN-based industrial Ethernet even in greenfield deployments. 
Therefore, the integration of 5G into existing industrial connectivity fabric 
would require devices to maintain synchronization with local time domains 
within a mixed 5G-Ethernet network as depicted in Fig.~\ref{fig:usecases}.
The transfer of precise timing reference (absolute or local domain time) via 
the 5G system to the devices using OTA synchronization procedure is currently 
being investigated in 3GPP Rel-16~\cite{3gpp.22.104}. 

Nevertheless, securing a robust distribution of reference time into the 
network has many associated challenges. Following the existing 
procedures in cellular systems, the devices must compensate for 
the propagation delay in the reference time. Typically, this is achieved using 
timing advance (TA)---the frame alignment procedure as used in LTE and the 5G 
NR radio interface. However, TA is an approximation of 
device-to-BS (D2B) propagation time since each TA value corresponds to a 
certain range of time of arrival (TOA) values. Both, the limited granularity 
of TA and the random perturbations in TOA due to the measurement and 
multipath errors, respectively, can introduce inaccuracy in reference time. 
%A possible solution to address TA related errors is to enable dedicated 
%signaling (same as PTP) but due to asymmetric uplink/downlink 
%propagation delays and nature of RRC scheduling makes it challenging.
Therefore, the impact of TA-related errors need a careful investigation and 
correction to meet the device-level synchronization target.

In the rest of this article, we present the role of time synchronization in 
multiple cMTC use cases in Section~\ref{sec:UseCases}. 
Section~\ref{sec:ExistingSolution} discusses the transition of 5G 
into industrial networks. Section~\ref{sec:OTA} presents the 
5G NR procedures and the associated timing errors to enable device-level 
synchronization. In Section~\ref{sec:Prop}, we quantify the errors in propagation delay due to 
TA and propose an improvement. Outline of new research 
directions to enhance synchronization accuracy in 5G concludes the paper.

%%%%%%%%%%%%%%%%%%%%%%%%%%%%%%%%%%%%%%%%%%%%%%%%%%%%%%%%%%%%%%%%%%%%%
%%%%%%%%%%%%%%%%%%%%%%%%%%%%%%%%%%%%%%%%%%%%%%%%%%%%%%%%%%%%%%%%%%%%%
\section{Use Cases of Device-Level Time Synchronization}
\label{sec:UseCases}

Any application requiring imperceptible lag in executing 
orders or reporting remote events 
%, with the devices operating in cohesion or 
%reporting correlated events, 
are the potential uses cases of device-level 
time synchronization. Albeit 5G-cMTC outlines several URLLC cases in 
\cite{3gpp.22.804}, we now elaborate the ones with device-level synchronization requisites 
(see Fig.~\ref{fig:usecases}).    
%ref: TS22.261. and 22.804\\

\subsection{Factory Automation}
\label{subsec:IndustrialAutomation}

\subsubsection{Isochronous real-time communication}
Critical industrial automation applications exhibit stringent requirement on 
communication latency and reliability, as well as on time-synchronized 
coordination among machines and robots. In particular, in closed-loop motion 
control---in packaging, printing and symmetrical welding/polishing---machines 
execute meticulously sequenced real-time tasks isochronously. Typically, a 
sequence of real-time control command/response frames is communicated over 
the communication links between controller and devices. To ensure smooth and 
deterministic execution of the production cycle, timely coordination among 
devices/machines must be accomplished, which is possible only if the 
devices are synchronized to a common time reference with clock disparity of less than 
\SI{1}{\micro\second}~\cite{3gpp.22.104}. 

\subsubsection{Data fusion}
Robust and accurate synchronization is required for meaningful sensor fusion, 
post-processing, and network analytics. In low-latency time-sensitive 
applications, the responses from various sensors must be fused to 
drive the logic behind the control systems. Such data fusion requires 
a synchronized collection of the events. Besides, the industrial IoT is 
meant to increase the operational efficiency based on data analytics of 
sensory information; however, the time context of sensor events must be 
factored in to make analytics definitive.

\subsection{Power Distribution Networks}
\label{subsec:PowerDistribution}

Managing power distribution networks---with an increasing amount of 
distributed energy resources and an increasing need of reliability, 
efficiency, and flexibility---requires enhanced communication 
technologies/services for functions as protection, control and remote 
monitoring~\cite{ABB5GSG}.
 
\subsubsection{Fault protection} 
In a transmission/distribution line, usually a line differential 
protection system detects faults based on a periodic sampling of electric 
current between the two relay devices~\cite{ABB5GSG}. When the relays differ 
in measurements, the system triggers the fault protection procedure, i.e., 
it sends a trip command to the relevant breaker. For such	 fault detection 
procedure to work correctly, 
relays are synchronized with an accuracy of \SI{< 20}{\micro\second}.     

\subsubsection{Control and optimization} 

With the increasing penetration of renewable resources, control and 
optimization are required at both the transmission and distribution level. A 
vital control task is to match power supply and demand as per the voltage and 
frequency regulations, where a control strategy, either centralized or 
distributed, is devised based on fine-grained information of measured 
electrical values of the load and the source. Although reliability and 
latency of reporting such information are specific to a control strategy, time 
synchronization accuracy is less demanding.

\subsubsection{Monitoring and diagnostics} 

Situational awareness and insights on the condition of distribution systems 
depend on measurements and analytics. There is a growing interest in 
instrumenting power distribution systems with phase measurement units 
(PMU)-like technology. PMUs take time-synchronized measurements of voltage and 
current to activate monitoring system for fault diagnosis. For instance, by 
synchronously measuring electric wave generated by a fault location at two 
points along the distribution line, the monitoring system can estimate the 
fault location based on the reported timing information. Precision 
in physical measurements, time synchronization accuracy, and the ability to 
cross-reference event locations offer more in-depth insights into the distribution 
network.  Usually, synchronization accuracy of higher than \SI{1}{\micro\second}
 is required to keep fault location uncertainty below 
\SI{300}{\meter}~\cite{Aamir_OTA}. 

\subsection{Vehicular Communication}

Vehicular communication is about wireless transactions among nearby vehicles 
and infrastructure---vehicle to everything (V2X) communication---that can 
enable cooperative intelligent transport systems capable of reducing traffic 
congestion and improving road safety. Ensuring timely and robust delivery of safety-related messages is 
critical. The examples of safety-related operations are forward collision 
warning and emergency electronic brake 
lights, which alert the drivers about possible collisions ahead. To execute 
these operations, vehicles need to exchange position, event 
timestamp, brake status, and heading information from onboard GPS. 
%To ensure the delivery of safety critical information, the clocks of the 
%vehicles/infrastructure/pedestrian must be synchronized. However,  
Furthermore, to function such coordination as per the age of information, the 
applications require that all the involved 
vehicles are synchronized to the same reference clock, usually 
provided by GPS. However, for such critical operations, a backup time 
synchronization service must be provided by the roadside communication infrastructure to 
handle GPS signal blockages and outages~\cite{HASAN_V2V}.

\subsection{Wireless Live Audio/Video Production}
\label{subsec:AV}

Wireless audio/video (A/V) equipment used for real-time production of 
audio-visual information, be it in the entertainment industry or live events and 
conferences, are denoted by the term program making and special events (PMSE). 
Usually, the wireless A/V production equipment includes cameras, microphones, 
in-ear monitors (IEM), conference systems, and mixing consoles. PMSE use 
cases are diverse, while each commonly being used for a limited duration in a 
confined local geographical area. In terms of communication requirements of 
typical live audio/video production setups, low-latency and ultra-reliable 
transmissions are pivotal to avoid failures and perceptible corruption of the 
media content. Moreover, perfect synchronization is crucial to minimize 
jitter among captured samples by multiple devices to render audio-video 
content~\cite{PMSE}.

For instance, in a demanding live audio performance, the microphone signal is 
streamed over a wireless channel to an audio mixing console where different 
incoming audio streams are mixed, and the in-ear audio mixes are streamed 
back to the microphone users via the wireless IEM system. For this, the audio 
sampling of microphones' signals must be perfectly synchronized to the system 
clock, which is usually integrated into the mixing console used for 
capturing, mixing, and playback of the audio signals. For immersive 3D audio 
effects, devices required synchronization accuracy of \SI{1}{\micro\second}, 
%with respect to reference clock 
which is higher than the audio 
sampling clock~\cite{PMSE,3gpp.22.804}.

%%%%%%%%%%%%%%%%%%%%%%%%%%%%%%%%%%%%%%%%%%%%%%%%%%%%%%%%%%%%%%%%%%%%%
%%%%%%%%%%%%%%%%%%%%%%%%%%%%%%%%%%%%%%%%%%%%%%%%%%%%%%%%%%%%%%%%%%%%%
\section{Keeping Devices in Sync: Existing Solutions and Emerging Requirements}
\label{sec:ExistingSolution}

The above-studied use cases manifest that embedded shared understanding of 
time at devices is essential for cMTC applications. To use 5G therein, a 
synchronization solution over the 5G system is a key URLLC enabler. Apart from 
delivering the desired accuracy, it should support the scenarios ranging from 
standalone operation to integration with existing/emerging solutions. 

\subsection{Existing Industrial Networks}
\subsubsection{Fragmented legacy solutions} in factory automation, the 
field devices---industrial devices and controller---are connected by various wired 
fieldbus and real-time Ethernet networks such as PROFIBUS, PROFINET, 
EtherCAT, Sercos and Modbus. While in power systems, the IEC 61850 series of 
standards specify networks for substation automation with profiles such as 
generic object-oriented substation event (GOOSE) and sampled values (SV).  
GOOSE is used for exchanging status, measurements, and interlocking signals 
between intelligent electrical devices (IEDs) while SV is used to transmit 
periodically sampled voltage and current measurements from measuring devices 
to IEDs~\cite{WirelessHP}. The wireless solutions (e.g., WirelessHART, 
ISA 100.11a, WIA-PA, WIA-FA) constitute only a small fraction of the 
installed base, and are used for non-critical connectivity of sensors over 
unlicensed bands.

\subsubsection{Time-sensitive networking (TSN)} the connectivity of 
industrial networks is expected to harmonize with the introduction of 
\textit{Ethernet with TSN support}---an open standard being developed by 
IEEE~802.1---where a TSN profile for 
industrial automation is being developed by the 
IEC/IEEE~60802~\cite{IEC-IEEE60802-TSN}. TSN 
includes the new features to standard Ethernet as \cite{TSN_Finn}: a) 
deterministic and bounded latency 
without congestion loss, b) priority queuing with resource allocation, c) 
reliability with redundant flows, and d) time synchronization among devices.

\subsubsection{Precision time protocol} pertaining to the transport of 
precise time and synchronization in industrial applications, 
variants (profiles) of IEEE 1588 (PTP) protocol are used. For example, PTP profile in 
TSN is IEEE 802.1AS while the synchronization profile in IEC 61850 is 
IEC/IEEE~61850-9-3. However, synchronization is kept separable from the rest of 
networking stack, thus reliability and timeliness features are independent of 
any particular synchronization protocol. 

\definecolor{Gray}{gray}{0.9}
\definecolor{Gray1}{gray}{0.7}
\bgroup
\def\arraystretch{1.2}
\begin{table*}[t]
	\caption{Timing Errors Associated to Delivery of Reference Time from BS to Devices~\cite{3gpp.TdocR1_1901252}}
	\centering
		\begin{tabular}{|p{0.23\linewidth}|p{0.47\linewidth}|p{0.20\linewidth}|}
			\hline
			%\rowcolor{Gray1}
		\textbf{Timing error source} &  \textbf{Description} &  \textbf{Typical 
values for SCS \newline [15 30 60 120] \SI{}{\kilo\hertz}} \\
		\hline
		\rowcolor{Gray}
		\multicolumn{3}{|l|}{\textbf{\textit{Reference time indication errors}}}\\
		\hline
		Time alignment error & Refers to desired synchronization accuracy among 
BSs for perfect frame timing required by new NR technologies and services. 
(see Section~\ref{subsec:BSasMaster}) & \hfil Tx diversity: $\sim$\SI{65}{
\nano\second} \newline \hspace*{23pt}
 Positioning: $\sim$
\SI{10}{\nano\second}\\
		\hline
		Reference time granularity& SIB16 granularity to transport reference time information. & \hfil
 \SI{250}{\nano\second}\\
		\hline
				UE DL frame timing estimation & Detection error of DL signal at 
UE, and the device's processing jitter & \hfil [390 260 227 114] ns\\
				\hline
		\rowcolor{Gray}
		\multicolumn{3}{|l|}{\textbf{\textit{TA related errors}}}
\\
		\hline 
		TA estimation error & TOA estimation is perturbed by measurement noise 
and multipath errors depending on the signal bandwidth and SNR of the direct 
path (Section~\ref{sec:Prop}). & \hfil Environment dependent \\
		\hline
		TA granularity & Error introduced by limited TA granularity, $ \pm 8 \cdot
 16 \cdot T_c /2^\mu$ (Section~\ref{subsec:TA}) & \hfil [260 130 65 32.5]
 ns\\
		\hline 
		TA adjustment error & The error at UE comprising
systematic and dynamic factors. & \hfil [130 130 65 16] ns\\
		\hline 
		Asymmetric DL/UL propagation \newline delay  & TA estimates UL 
propagation delay while reference time indication needs adjustment with DL 
propagation delay. Asymmetry in DL/UL propagation delay (FDD) will 
introduce inaccuracy in TA-compensated reference time. & \hfil Negligible in 
TDD\\
		\hline 
		UE UL transmit timing error  & The jitter in UL transmit time contributes 
to TOA estimation error, which could be considered to be negated by DL frame timing 
error. & Same as "UE DL frame timing estimation error" \\
		\hline 
		\rowcolor{Gray}
		\multicolumn{3}{|l|}{\textbf{\textit{Other errors}}}\\
		\hline
		UE modem to host interface \newline chipset delay & Delay introduced by 
the 
interface between the device modem and the host chipset maintaining clock 
information. & 
\hfil $\sim$\SI{65}{
\nano\second}\\
		\hline 
		\end{tabular}
	\label{tab:TimingErrors}
	\vspace{-7pt}
\end{table*}
\egroup

\subsection{5G Synchronization Requirements}

\subsubsection{Timing service in RAN} GPS could provide an accurate but 
costly solution to establish UTC time-reference at devices. Further, jamming 
and weak signal reception raise concerns in indoor deployments. In indoor 
deployments, although a GPS antenna could be installed 
outdoors to enhance signal reception, long feeder cable with an amplifier 
from the antenna to the receiver is required, which is costly and inflexible. Consequently, there is an interest in built-in timing service 
over cellular networks. The 5G network can be considered stable and scalable; 
however, there is a need to upgrade the 5G air interface to distribute accurate 
time reference to the devices.

\subsubsection{Unified 5G-TSN network}

5G is expected to satisfy most of the cMTC applications with new RAN features 
like faster scheduling, short/robust transmissions, faster retransmissions, 
preemption and packet duplication, as well as diversity techniques. However, 
it will replace the existing systems in multiple phases, primarily driven by 
the benefits (cost, capabilities) of introducing 5G connectivity. Even in 
greenfield industrial deployments, not all industrial networks will be 
migrated to 5G. Therein, the 5G local industrial network will coexist with 
traditional networks and might even require transparent integration to transport 
industrial Ethernet or TSN. In such scenarios, collaborative actions 
of devices belonging to different domains need to be coordinated in time. 
Accordingly, the 5G system will need to relate/synchronize devices to a 
master clock of one or more time domains to enable time-scheduled 
coordination over a combined 5G-TSN network~\cite{5GSmartManufactu}.

%%%%%%%%%%%%%%%%%%%%%%%%%%%%%%%%%%%%%%%%%%%%%%%%%%%%%%%%%%%%%%%%%%%%%
%%%%%%%%%%%%%%%%%%%%%%%%%%%%%%%%%%%%%%%%%%%%%%%%%%%%%%%%%%%%%%%%%%%%%
\section{Synchronization in 5G RAN}
\label{sec:OTA}

In this section, we study the 5G radio interface to transport reference time 
from BS to the devices. Usually, a common notion of time can be maintained by 
periodically broadcasting timestamps of reference time from master (i.e., BS) 
to the slave devices. The devices use timestamp information to align their clocks after removing 
any time progress from the timestamping-to-reception instance of the 
message~\cite{Aamir_OTA}. The synchronization period depends on the frequency and phase 
stability of the onboard oscillators in devices, causing clock skew and drift. 
In this process, the main disrupting elements to synchronization accuracy are:  
\begin{itemize}[leftmargin=*]
	\item[--] \textit{BS related}: time alignment error (TAE), timestamping to transmission delay, and 
timestamping granularity.
	\item[--] \textit{Channel related}: propagation time and its variations (jitter), asymmetry uplink/
downlink propagation, and scheduling/medium access delays.
	\item[--] \textit{Device related}: time adjustment errors at device, 5G device to IIoT host 
interface delay.
\end{itemize}
In reference to a synchronization procedure currently being investigated in 
3GPP release-16, we discuss in the following, (a) possible TAE at BS, (b)  
reference time indication procedure from BS to devices, and (c) propagation 
delay adjustment in reference time. Table~\ref{tab:TimingErrors} summarizes 
the timing errors associated with the synchronization steps (b) inclusive of 
(a) and (c). 

\subsection{Time Alignment Error at BS}
\label{subsec:BSasMaster}

Any TAE at the BS will add up in time uncertainty at the devices. The TAE 
requirements for new 5G technologies/services are summarized as 
follows~\cite{TS_5G}.
\begin{itemize}[leftmargin=*]
	\item \textit{T\text{\normalfont x} diversity}: TAE requirement for MIMO 
and Tx diversity is \SI{+-65}{\nano\second}.
	\item \textit{Carrier aggregation (CA)}: enables the use of multiple 
contiguous or non-contiguous intra-band/inter-band carriers to increase 
throughput. CA can be performed at intra-BS and inter-BS level where TAE for 
inter-BS CA must be  \SI{<260}{\nano\second}.
  	\item \textit{Coordinated multi-point (CoMP)}: joint transmission from 
multiple BSs requires time offset at a device within 
$[-0.5, 2]~$\SI{}{\micro\second}, which includes TAE and the difference in 
propagation delays 
where inter-site $\mathrm{TAE}\leq$\SI{260}{\nano\second} is required. 
  	\item \textit{New frame structure}: to avoid overlap in uplink and 
downlink timeslots in TDD systems, new 5G frame structure requires accuracy 
of \SI{+-390}{\nano\second} in D2B alignment.   
  	\item \textit{Positioning}: 5G positioning services seeking \SI{3}{\meter} 
location accuracy require $\mathrm{TAE}$ of \SI{+-10}{\nano\second} among BSs.
\end{itemize}

\subsection{Reference Time Indication (RTI)}

RTI is concerned with the distribution of reference time in 5G RAN, from BS 
to devices. The reference time could be either the 5G or TSN network's clock. %, which depends on the clock distribution model. 
To distribute the 5G network clock as a reference, 3GPP considers 5G radio 
interface signaling as dedicated RRC or system information block (SIB) 
broadcasts~\cite{3gpp.TdocR2_1817172}. Whereas, to support the distribution of 
TSN clock, the 5G system acts as an IEEE~802.1AS time-aware system by adding 
TSN translators (TTs) in the wireless edge, i.e., before a BS and after each 
device~\cite{3gpp.Tdoc_Joa}. Only TTs support the IEEE~802.1AS operations; particularly, the 
timestamping of a PTP sync message at TTs using 5G clock and forwarding of 
the PTP sync via user plane PDU. The difference in egress and ingress 
timestamps of a PTP sync message determines the \textit{residence time} in 
the 5G system. 

Hence, distribution of the 5G system clock is a prerequisite to establish any 
reference time at devices, while the following aspects of SIB/RRC 
signaling can introduce time uncertainty:
\begin{itemize}[leftmargin=*]
\item[] \textit{Time progression adjustment:} BS needs to adjust the acquired 
reference time with a projected time of transmission: that is, up to a 
reference point in RTI frame occurring at the antenna reference point. 
	\item[] \textit{Granularity:} timestamping granularity of SIB messages could 
introduce time uncertainty of up to \SI{250}{\nano\second}.     
\end{itemize}

\subsection{Propagation Delay Compensation}
\label{subsec:TA}
Accurate estimation of propagation delay is a key factor to enable 
device-level synchronization. It is required to adjust the time progression 
after SIB timestamping/transmission. However, the need for propagation delay 
compensation depends on the service area. 
In the case of a small area (e.g., \SI{10}{\meter\squared}) the propagation delay is almost 
negligible (i.e., \SI{0.3}{\nano\second}), and the timing inaccuracy is the sum of reference 
time indication errors as listed in Table~\ref{tab:TimingErrors}. 
For larger areas, 3GPP resort to utilizing timing advance (TA)  in 
combination with SIB16. In LTE and 5G systems, TA is used to adjust the 
uplink transmission time of 
the devices based on their respective propagation delays in order
to avoid collisions at the BS. 

TA is negotiated during network access and RRC connected state using uplink 
reference signals: PRACH, SRS, and DMRS. During network access, BS estimates 
TA from \textit{network access request} from a device and  issues a TA 
command in \textit{random access response} with $N_{\textrm{TA}}$ values with 
index $T_A = 0,1,2,\cdots,3846$. The value of time alignment with subcarrier 
spacing (SCS) of $2^\mu\cdot$ \SI{15}{\kilo\hertz} is multiple of 
$T_\mu = 16 \cdot 64 \cdot T_c / 2^\mu$~sec before the start of the corresponding 
downlink frame. Here, $\mu=0,1,2,\cdots$ defines the NR numerology, 
$T_c = 1/(\Delta f_{\max} \cdot N_f)$ is the 5G basic time unit with 
$\Delta f_{\max} =$ \SI{480}{\kilo\hertz} and $N_f=4096$. In RRC connected state, TA is 
negotiated with periodic control messages to adjust the uplink timing 
relative to current timing. The time alignment has the index value 
$T_A = 0,1,\cdots, 63$, which adjusts the current uplink timing by 
$(T_A-31) \cdot T_\mu$~sec. 

As mentioned earlier, a device can use TA as an approximation of TOA---the 
propagation time from the device to BS---in the absence of original TOA 
measurement. As each TA corresponds to a range of TOA measurements 
with timeslot $T = T_\mu/2$ of limited granularity, TA can yield a 
maximum synchronization error of $\pm T = 8 \cdot 16 \cdot T_c/2^\mu$. 
Moreover, the random errors in the original TOA could lead to wrong TA 
selection. The impact of random errors on TOA estimation from TA is further 
discussed in Section~\ref{sec:Prop}.

\subsection{Other Timing Errors}

\begin{figure*}[!t] 	
	\centering 		
	\includegraphics[width=0.75\linewidth]{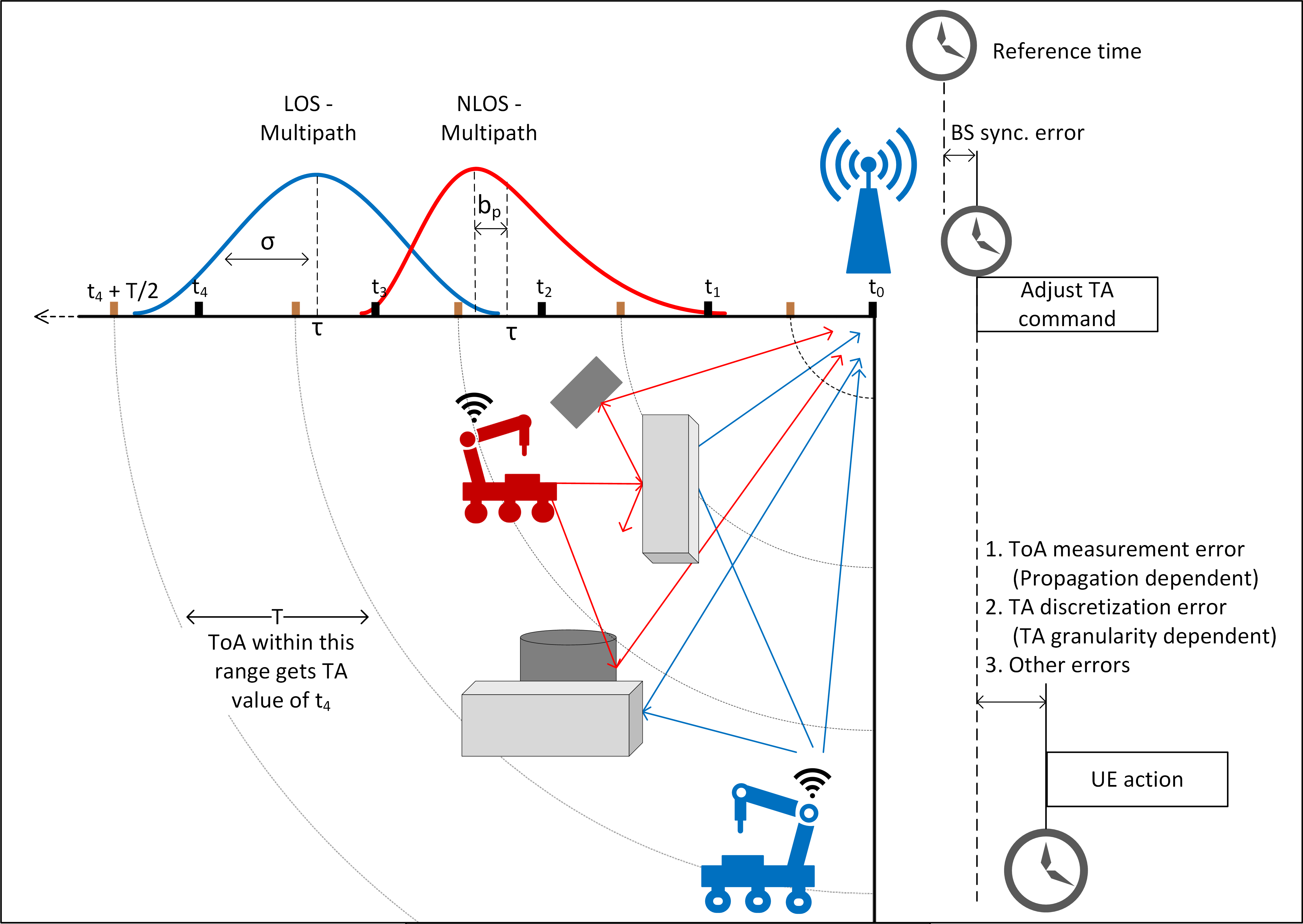}
 	\caption{Transport of reference time information to the devices: Principles and error components.}
	\label{fig:TA} 
	\vspace{-10pt}
\end{figure*}

The above-discussed radio parameters have other associated timing errors, 
which must be considered in BS-to-device time offset budget to find synchronization 
inaccuracy. The components that could impact the accuracy are elaborated in 
Table~\ref{tab:TimingErrors} and can be logically visualized in Fig.~\ref{fig:TA}.

\section{Analysis of Error in Time of Arrival}
\label{sec:Prop}
We analyze the error in estimating TOA from TA while considering measurement 
errors in true TOA. A BS measures the unknown TOA $\tau$ of a radio 
signal with a certain random error $\sigma$, which is a function of signal 
bandwidth and signal-to-noise ratio (SNR)~\cite{gentile2012geolocation}. For 
TOA values in a timeslot with width $T$ and center at $t_i$, the BS assigns 
the TA bin $i$ as illustrated in Fig.~\ref{fig:TA}. If $\tau$ falls within 
one or two $\sigma$ of timeslot borders $t_i \pm T/2$, then there is a 
non-negligible probability that a wrong TA bin is selected. Selection of a 
wrong 
timeslot adds at least $T/2$ to the error in $\tau$. Thus, both the random 
error $\sigma$ and the timeslot width $T$ must be considered to find the 
value of $\tau$ from TA.

%%%%%%%%%%%%%%%%%%%%%%%%%%%%%%%%%%%%%%%%%%%%%%%%%%%%%%%%%%%%%%%%%%%%%
%%%%%%%%%%%%%%%%%%%%%%%%%%%%%%%%%%%%%%%%%%%%%%%%%%%%%%%%%%%%%%%%%%%%%
%\section{Analysis of TA with 5G Numerologies}

\begin{figure}[!t] 	
	\centering 		
	\includegraphics[width=1\linewidth]{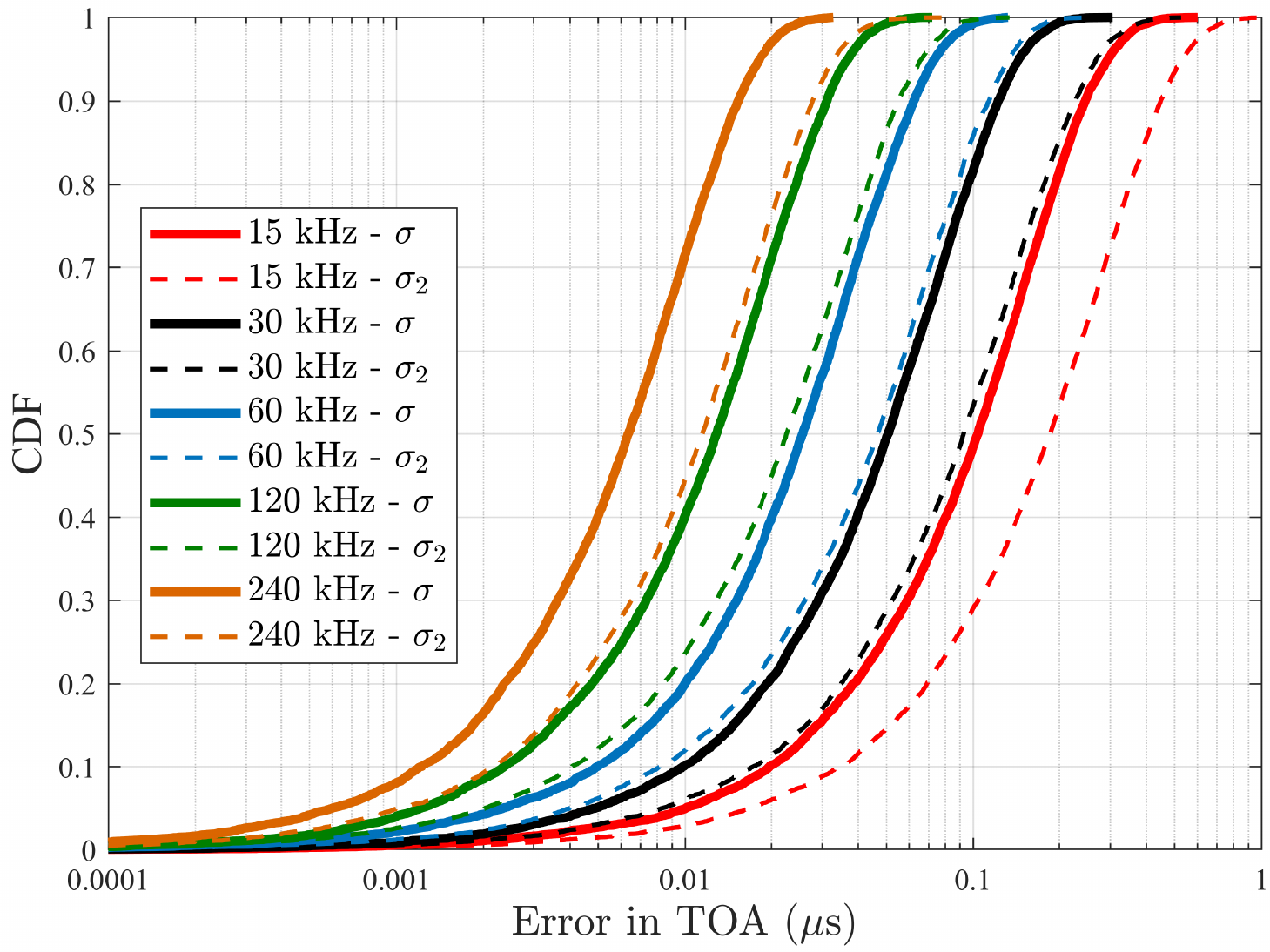}
 	\caption{CDFs of the difference between true TOA and estimated TOA from 
reported TA, where $\sigma$ is the variance of measurement error in true TOA.} 	
	\label{fig:num} 
	\vspace{-10pt}
\end{figure}

We studied the error in true TOA and the estimated TOA extracted from reported 
TA by simulations. We assume a uniform distribution of true TOA in 
timeslot $t_i \pm T/2$, where true TOA is perturbed by Gaussian errors with 
$\sigma = T/2$ and $\sigma_2 = T$ but without any bias. It implies that the $\sigma$ 
reduces with the increase in SCS; a receiver's ability to resolve TOA of 
multipath signal components improves with increase in subcarrier spacing. 
Fig.~\ref{fig:num} shows the cumulative distribution functions (CDFs) of the studied error for 
different SCSs. The CDF curves are useful to define the confidence level in 
synchronization accuracy when adjusting the propagation delay in reference 
time indication. The point on the $x$-axis with $P(X \leq x) = 1$ 
guarantees that the synchronization error introduced by TA is less than equal 
to $x$ with 100\% confidence, and must be used in the device synchronization 
budget. If we denote this point as $P_e$, it can be observed that as SCS 
increases the $P_e$ reduces.
%, while the increase in random error 
%increases $P_e$ by a factor of $T/2$ (check this). 
Note that if the timing errors in Table~\ref{tab:TimingErrors} are taken into 
account, $P_e$ for SCS \SI{15}{\kilo\hertz} is high enough not to satisfy 
\SI{1}{\micro\second} target.

\begin{figure}[!t] 	
	\centering 		
	\includegraphics[width=1\linewidth]{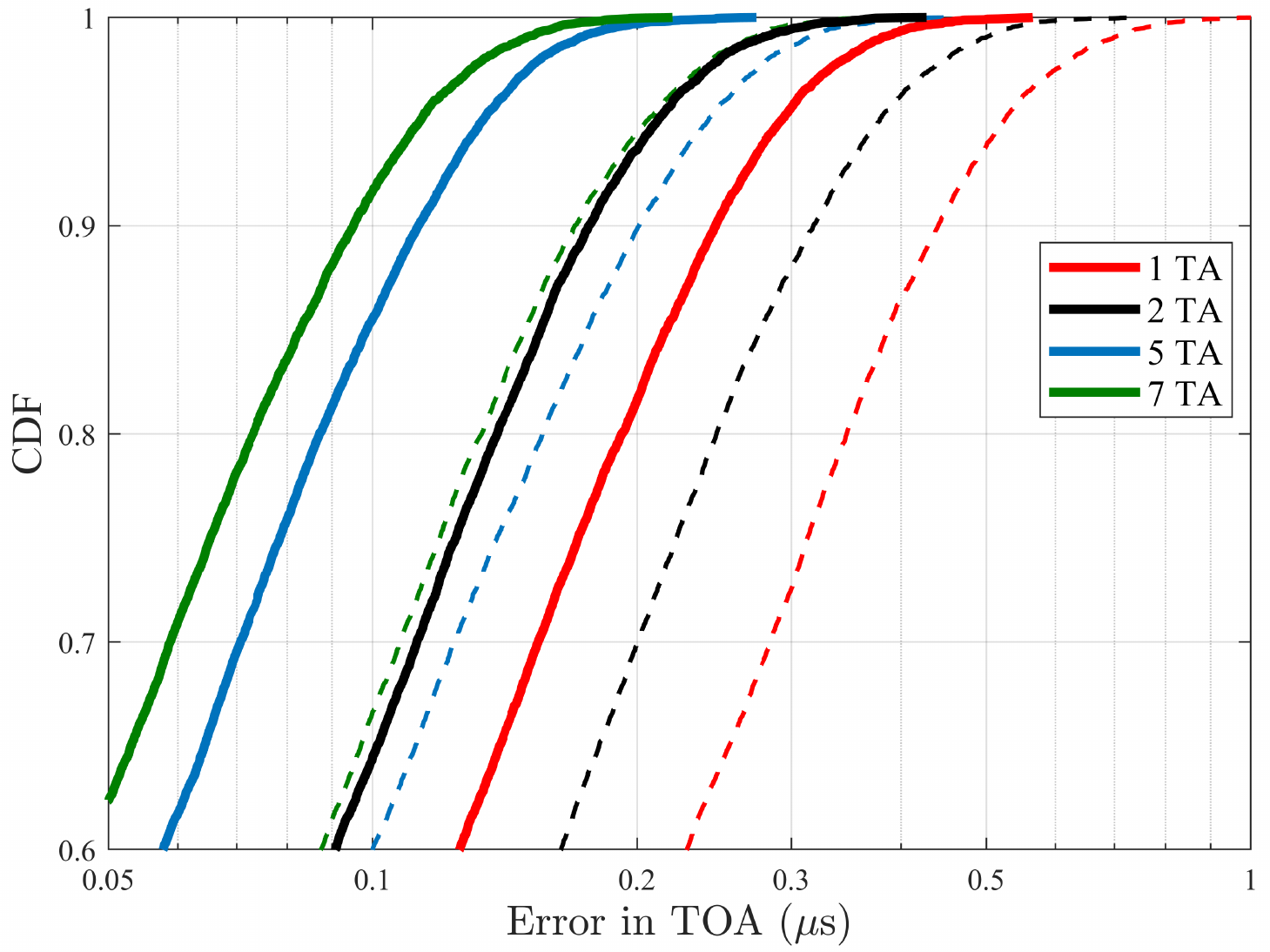}
 	\caption{Error reduction by averaging multiple TAs for \SI{15}{\kilo\hertz} 
subcarrier spacing. Solid curves: $\sigma=T/2$, dashed curves: $\sigma=T$.}
	\label{fig:TA_avg} 
	\vspace{-10pt}
\end{figure}

One solution to reduce the error $P_e$ is to take an average of two or more 
consecutive TAs. The averaging reduces the error caused by the TOA 
measurements that are assigned TA values to the sides of true TA. 
For SCS \SI{15}{\kilo\hertz}, Fig.~\ref{fig:TA_avg} shows that error 
reduction obtained by averaging multiple consecutive TAs is substantial. 
Therefore, a required synchronization accuracy target can be 
achieved by appropriate selection of averaging size for a given measurement 
random error, which is influenced by the propagation conditions.

\textbf{Random error under multipath fading}: Fig.~\ref{fig:num} shows that 
TOA estimation from TA depends on the amount of perturbation in true TOA. In 
densely cluttered environments, the true TOA is perturbed by both the 
measurement noise and LOS/NLOS multipath error. In LOS multipath 
environments, multipath signals tend to arrive close to the direct 
path. The signals combine to create a cluster in power delay 
profile, making it challenging to extract TOA of the direct path. As a result, 
depending on the structure of the propagation environment, TOA estimation 
from TA may lead to varying synchronization errors. The statistics of LOS 
multipath errors can be modeled as a zero-mean Gaussian with variance 
directly related to the variations in the multipath 
structure~\cite{gentile2012geolocation}.

Contrarily, NLOS multipath environment is challenging because of multipath 
errors, where the TOA estimation depends on the detection of direct path (DP). If 
the attenuated DP is detectable (consider light obstructions), better TOA 
estimation can be achieved. On the other hand, in case the DP is buried in 
noise, it will create a bias ($b_p$ in Fig.~\ref{fig:TA}) towards a longer first non-DP. Since shadowing 
introduces fluctuations in the detection of the first arrival path, the 
variance of multipath error is also time varying. Clearly, NLOS 
introduces bias as well as other perturbations in TOA estimation. One 
technique to remove bias could be to introduce some correction for it. The 
other issue is the asymmetric distribution of random errors that may require 
a TOA estimator other than the timeslot center.

%%%%%%%%%%%%%%%%%%%%%%%%%%%%%%%%%%%%%%%%%%%%%%%%%%%%%%%%%%%%%%%%%%%%%
%%%%%%%%%%%%%%%%%%%%%%%%%%%%%%%%%%%%%%%%%%%%%%%%%%%%%%%%%%%%%%%%%%%%%
%%%%%%%%%%%%%%%%%%%%%%%%%%%%%%%%%%%%%%%%%%%%%%%%%%%%%%%%%%%%%%%%%%%%%
%%%%%%%%%%%%%%%%%%%%%%%%%%%%%%%%%%%%%%%%%%%%%%%%%%%%%%%%%%%%%%%%%%%%%
%%%%%%%%%%%%%%%%%%%%%%%%%%%%%%%%%%%%%%%%%%%%%%%%%%%%%%%%%%%%%%%%%%%%%
%%%%%%%%%%%%%%%%%%%%%%%%%%%%%%%%%%%%%%%%%%%%%%%%%%%%%%%%%%%%%%%%%%%%%
%%%%%%%%%%%%%%%%%%%%%%%%%%%%%%%%%%%%%%%%%%%%%%%%%%%%%%%%%%%%%%%%%%%%%
%%%%%%%%%%%%%%%%%%%%%%%%%%%%%%%%%%%%%%%%%%%%%%%%%%%%%%%%%%%%%%%%%%%%%
%%%%%%%%%%%%%%%%%%%%%%%%%%%%%%%%%%%%%%%%%%%%%%%%%%%%%%%%%%%%%%%%%%%%%
%%%%%%%%%%%%%%%%%%%%%%%%%%%%%%%%%%%%%%%%%%%%%%%%%%%%%%%%%%%%%%%%%%%%%
%%%%%%%%%%%%%%%%%%%%%%%%%%%%%%%%%%%%%%%%%%%%%%%%%%%%%%%%%%%%%%%%%%%%%
%%%%%%%%%%%%%%%%%%%%%%%%%%%%%%%%%%%%%%%%%%%%%%%%%%%%%%%%%%%%%%%%%%%%%
\section{Conclusions and Research Directions}
Sharing a common time-base among devices is essential for cMTC applications 
to perform various tasks; ranging from coordination, sampling and fusion, and 
event reconstruction. Together with low-latency and ultra-reliability, 
enabling ultra-tight time synchronization can be regarded as the 
third dimension of 5G RAN enhancements. To operate either in standalone or in 
cohesion with TSN/Ethernet solutions, transport of reference time over the 5G air 
interface is currently being investigated in 3GPP release 16 in order to 
enable device-level synchronization across multiple domains. In this paper, 
we discussed enabling radio parameters in 5G NR and focused on propagation 
time compensation in reference time based on timing advance (TA). Timing 
advance corresponds to a set of TOA values, which is perturbed by signal 
propagation conditions, and could lead to substantial errors in time 
synchronization. We studied the TA-dependent timing error and observed that 
the averaging of multiple TAs could reduce the error and satisfy the overall 
accuracy target. Nevertheless, there are still many research areas to be 
addressed, for instance: i) impact of  mobility on TA averaging, ii) TOA 
uplink and downlink asymmetry, iii) bias and TOA error asymmetry in NLOS 
conditions.   

% use section* for acknowledgment
%%%%%%\section*{Acknowledgment}
%%%%%%The authors would like to thank...

% Can use something like this to put references on a page
% by themselves when using endfloat and the captionsoff option.
%%%%%\ifCLASSOPTIONcaptionsoff
  %%%%%\newpage
%%%%%\fi

% trigger a \newpage just before the given reference
% number - used to balance the columns on the last page
% adjust value as needed - may need to be readjusted if
% the document is modified later
%\IEEEtriggeratref{8}
% The "triggered" command can be changed if desired:
%\IEEEtriggercmd{\enlargethispage{-5in}}

% references section

% can use a bibliography generated by BibTeX as a .bbl file
% BibTeX documentation can be easily obtained at:
% http://mirror.ctan.org/biblio/bibtex/contrib/doc/
% The IEEEtran BibTeX style support page is at:
% http://www.michaelshell.org/tex/ieeetran/bibtex/
%\bibliographystyle{IEEEtran}
% argument is your BibTeX string definitions and bibliography database(s)
%\bibliography{IEEEabrv,../bib/paper}
%
% <OR> manually copy in the resultant .bbl file
% set second argument of \begin to the number of references
% (used to reserve space for the reference number labels box)

\bibliographystyle{IEEEtran}
\bibliography{TSinURLLC1}

%%%%%%%%%%%%%%%%%%%%%%%%%%%%%%%%%%%%%%%%%%%%%%%%%%%%%%%%%%%%%%%%%%%%%
%%%%%%%%%%%%%%%%%%%%%%%%%%%%%%%%%%%%%%%%%%%%%%%%%%%%%%%%%%%%%%%%%%%%%
%%%%%%%%%%%%%%%%%%%%%%%%%%%%%%%%%%%%%%%%%%%%%%%%%%%%%%%%%%%%%%%%%%%%%
%%%%%%%%%%%%%%%%%%%%%%%%%%%%%%%%%%%%%%%%%%%%%%%%%%%%%%%%%%%%%%%%%%%%%
%%%%%%%%%%%%%%%%%%%%%%%%%%%%%%%%%%%%%%%%%%%%%%%%%%%%%%%%%%%%%%%%%%%%%
%%%%%%%%%%%%%%%%%%%%%%%biography section%%%%%%%%%%%%%%%%%%%%%%%%%%%%%
%%%%%%%%%%%%%%%%%%%%%%%%%%%%%%%%%%%%%%%%%%%%%%%%%%%%%%%%%%%%%%%%%%%%%
%%%%%%%%%%%%%%%%%%%%%%%%%%%%%%%%%%%%%%%%%%%%%%%%%%%%%%%%%%%%%%%%%%%%%
%%%%%%%%%%%%%%%%%%%%%%%%%%%%%%%%%%%%%%%%%%%%%%%%%%%%%%%%%%%%%%%%%%%%%
%%%%%%%%%%%%%%%%%%%%%%%%%%%%%%%%%%%%%%%%%%%%%%%%%%%%%%%%%%%%%%%%%%%%%
%%%%%%%%%%%%%%%%%%%%%%%%%%%%%%%%%%%%%%%%%%%%%%%%%%%%%%%%%%%%%%%%%%%%%
%%%%%%%%%%%%%%%%%%%%%%%%%%%%%%%%%%%%%%%%%%%%%%%%%%%%%%%%%%%%%%%%%%%%%
% If you have an EPS/PDF photo (graphicx package needed) extra braces are
% needed around the contents of the optional argument to biography to prevent
% the LaTeX parser from getting confused when it sees the complicated
% \includegraphics command within an optional argument. (You could create
% your own custom macro containing the \includegraphics command to make things
% simpler here.)
%\begin{IEEEbiography}[{\includegraphics[width=1in,height=1.25in,clip,keepaspectratio]{mshell}}]{Michael Shell}
% or if you just want to reserve a space for a photo:
\vspace{-100pt}
\begin{IEEEbiographynophoto}{Aamir Mahmood}
[M'18] (aamir.mahmood@miun.se) is an assistant professor of communication 
engineering at Mid Sweden University, Sweden. He received the M.Sc. and D.Sc. 
degrees in communications engineering from Aalto University School of 
Electrical Engineering, Finland, in 2008 and 2014, respectively. He was a 
research intern at Nokia Researcher Center, Finland and a visiting researcher 
at Aalto University during 2014-2016. He has served as publications chair of 
the IEEE WFCS 2019. His research interests include network time 
synchronization, resource allocation for URLLC, and 
RF interference/coexistence management.
\end{IEEEbiographynophoto}
\vspace{-120pt}
\begin{IEEEbiographynophoto}{Muhammad Ikram Ashraf}
(ikram.ashraf@ericsson.com) is an experienced researcher in network 
architecture and protocols at Ericsson Research in Jorvas, Finland. He joined 
Ericsson in 2018 and mainly works on industrial wireless communication with a 
focus on URLLC. Before joining Ericsson, Ikram worked for Nokia Bell-Labs and 
Centre for Wireless Communication (CWC) at the University of Oulu, Finland. 
He has authored/co-authored more than 30 international scientific 
publications and has contributed in several standardization contributions. 
His current research interest includes 5G radio networks, Industrial IoT, 
TSN, and V2X communication.
\end{IEEEbiographynophoto}
\vspace{-120pt}
\begin{IEEEbiographynophoto}{Mikael Gidlund}
[M'98--SM'16] (mikael.gidlund@miun.se) is a professor of computer engineering 
at Mid Sweden University, Sweden. He has worked as Senior Principal Scientist 
and Global Research Area Coordinator of Wireless Technologies, ABB Corporate 
Research, Sweden, Project Manager and Senior Specialist with Nera Networks 
AS, Norway, and Research Engineer and Project Manager with Acreo AB, Sweden. 
His current research interests include wireless communication and networks, 
wireless sensor networks, access protocols, and security. He holds more than 
20 patents in the area of wireless communication. He is an associate editor 
of the IEEE Transactions on Industrial Informatics.
\end{IEEEbiographynophoto}
\vspace{-120pt}
\begin{IEEEbiographynophoto}{Johan Torsner}
(johan.torsner@ericsson.com) is a research manager at Ericsson Research and 
is currently leading Ericsson's research activities in Finland. He joined 
Ericsson in 1998 and has held several positions within research and R\&D. He 
has been deeply involved in the development and standardization of 3G and 4G 
systems and has filed over 100 patent applications. His current research 
interests include 4G evolution, 5G, and machine-type communication. 
\end{IEEEbiographynophoto}
\vspace{-120pt}
\begin{IEEEbiographynophoto}{Joachim Sachs}
(joachim.sachs@ericsson.com) is a principal researcher at 
Ericsson Research. He joined Ericsson Research in 1997 and currently holds 
the position of principal researcher in network architectures and protocols. 
He has contributed to the development and evaluation of 3G and 4G, and today 
coordinates the research and standardization activities on machine-type 
communication and URLLC for 5G. He holds a diploma in electrical engineering 
from Aachen University (RWTH), and a doctorate in electrical engineering from 
the Technical University of Berlin, Germany. Since 1995, he has been active 
in the IEEE and the German VDE Information Technology Society (ITG), where he 
is currently co-chair of the technical committee on communication.
\end{IEEEbiographynophoto}
% if you will not have a photo at all:
%%%%%\begin{IEEEbiographynophoto}{John Doe}
%%%%%Biography text here.
%%%%%\end{IEEEbiographynophoto}

% insert where needed to balance the two columns on the last page with
% biographies
%\newpage

%%%%%\begin{IEEEbiographynophoto}{Jane Doe}
%%%%%Biography text here.
%%%%%\end{IEEEbiographynophoto}

% You can push biographies down or up by placing
% a \vfill before or after them. The appropriate
% use of \vfill depends on what kind of text is
% on the last page and whether or not the columns
% are being equalized.

%\vfill

% Can be used to pull up biographies so that the bottom of the last one
% is flush with the other column.
%\enlargethispage{-5in}
\end{document}